\documentclass[aps,a4paper,10pt,twocolumn,nofootinbib]{revtex4} 
\usepackage{mathptmx}
\usepackage{amsmath}
\usepackage{amsfonts}
\usepackage{amsfonts}
\usepackage{eufrak}
\usepackage{mathrsfs}
\usepackage[mathscr]{euscript}
\usepackage[dvips]{graphicx}
\usepackage{fancyhdr}
\usepackage{colordvi}

\def\overstrike#1#2{{\setbox0\hbox{$#2$}\hbox to \wd0{\hss
    $#1$\hss}\kern-\wd0\box0}}

\newcommand{\cross}{\times}

        \DeclareMathOperator{\grad}{\nabla}

\begin{document}
\title{Four Poynting theorems}
\author{Paul Kinsler}
\email{Dr.Paul.Kinsler@physics.org}
\author{Alberto Favaro}
\author{Martin W. McCall}
\affiliation{
  Blackett Laboratory, Imperial College,
  Prince Consort Road,
  London SW7 2AZ, 
  United Kingdom.
}

\lhead{\includegraphics[height=5mm,angle=0]{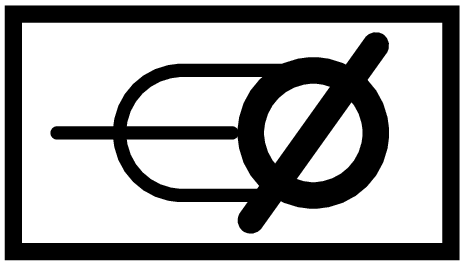}~~EMCON4}
\chead{Four Poynting theorems}
\rhead{
\href{mailto:Dr.Paul.Kinsler@physics.org}{Dr.Paul.Kinsler@physics.org}\\
\href{http://www.kinsler.org/physics/}{http://www.kinsler.org/physics/}
}

\begin{abstract}

The Poynting vector
 is an invaluable tool for analysing electromagnetic problems.
However,
 even a rigorous stress-energy tensor approach
 can still leave us with the question:
 is it best defined
 as $\Vec{E} \cross \Vec{H}$ or as $\Vec{D} \cross \Vec{B}$?
Typical electromagnetic treatments provide yet another perspective:
 they regard $\Vec{E} \cross \Vec{B}$ as the appropriate definition, 
 because $\Vec{E}$ and $\Vec{B}$
 are taken to be the fundamental electromagnetic fields.
The astute reader will even notice
 the fourth possible combination of fields:
 i.e. $\Vec{D} \cross \Vec{H}$.
Faced with this diverse selection,
 we have decided to treat each possible flux vector on its merits,
 deriving its associated energy continuity equation
 but applying minimal restrictions to the allowed host media.
We then discuss each form,
 and how it represents the response of the medium.
Finally,
 we derive a propagation equation for each flux vector
 using a directional fields approach;
 a useful result which enables further interpretation
 of each flux and its interaction with the medium.

\end{abstract}

\date{\today}
\maketitle
\thispagestyle{fancy}

\noindent{\emph{Published in Eur. J. Phys. \textbf{30}, 983 (2009).}\footnote{
 This is an author-created, un-copyedited version of an article
  accepted for publication in the European Journal of Physics. 
 IOP Publishing Ltd is not responsible for any errors or omissions
  in this version of the manuscript or any version derived from it. 
 The definitive publisher-authenticated version is available online
 at doi:10.1088/0143-0807/30/5/007.}
}\\
\small{This arXiv version has updates not present in the published version.}

%
\section{Introduction}\label{S-intro}

The correct definition of electromagnetic flux
 has long been controversial,
 with the main competition being between
 the Abraham $\Vec{E}\cross\Vec{H}$ \cite{Abraham-1909rcmp,Abraham-1910rcmp}
 and Minkowski $\Vec{D}\cross\Vec{B}$ \cite{Minkowski-1908} forms.
Pfeifer et al. \cite{Pfeifer-NHR-2007rmp} gave
 an excellent discussion and historical review of the situation,
 with an analysis based on energy-momentum tensors.
In contrast,
 the Poynting theorem \cite{Poynting-1884rspl} and Poynting vector
 $\Vec{S} = \Vec{E} \cross \Vec{H}$,
 in concert with the electromagnetic energy density,
 lead us to an energy continuity equation.
This equation is easily interpreted when
 considering only fields in the vacuum
 or in nondispersive linear media
 \cite{LandauLifshitz}.
Further alternatives exist where the Poynting vector is generalized
 to include extra terms and so generate other equally valid
 flux vectors and energy densities
 \cite{Slepian-1942jap,Lai-1981ajp,Peters-1982ajp,Romer-1982ajp,Kobe-1982ajp};
 however all these were based on $\Vec{E} \cross \Vec{H}$.

However,
 outside the context of energy-momentum tensor definitions,
 but nevertheless common in electromagnetic usage,
 is the $\Vec{E}\cross\Vec{B}$ form
 \cite{Feynman-Lectures-vII,MelroseMcPhedran-EMPDM}.
Curiously,
 comparison or contrast of the Abraham/ Minkowski forms
 with $\Vec{E}\cross\Vec{B}$ is hard to find --
 e.g. \cite{Pfeifer-NHR-2007rmp} does not remark
 on the different origins of $\Vec{E}\cross\Vec{B}$
 at all.
Further,
 even though (e.g.)
 both $\Vec{E}\cross\Vec{H}$ and $\Vec{E}\cross\Vec{B}$ 
 appear in \cite{RMC}, 
 this is outside the context of magnetic media, 
 so that $\Vec{H} \equiv \Vec{B}/\mu_0$ in any case.

In this paper we address the question:
 if we construct electromagnetic flux vectors (``Poynting vectors'')
 using the cross product of two fields, 
 what do the results look like, 
 and what might we apply them to?
How should non-trivial polarization and magnetization effects be interpreted?
Also,
 does the alternative $\Vec{D}\cross\Vec{H}$ combination of field vectors
 give interesting results?
We allow for reasonably general propagation media,
 with potentially dispersive and nonlinear properties
 affecting \emph{both} electric and magnetic fields.
Such complications mean that we do not derive
 continuity equations with a perfectly balanced 
 flux and energy density, 
 but they also include extra ``residual'' terms:
 e.g. the standard $\Vec{E}\cross\Vec{B}$ derivation
 produces a residual ``work done'' term $\Vec{E} \cdot \Vec{J}$, 
 where $\Vec{J}$ is the total current
 \cite{MelroseMcPhedran-EMPDM,Richter-FH-2008epl}.
We show that in each case the residuals
 contain either temporal or spatial derivatives, 
 and that they can be interpreted in terms of currents; 
 we further show how 
 they affect the propagation of the flux vector.

Since typical response models for a propagation medium, 
 especially in the nonlinear case,
 are likely to be non-covariant, 
 we tie our description to the medium rest frame
 without further loss of generality.
Although this is a restriction we might prefer to avoid,
 as done for negative refraction by McCall \cite{McCall-2008meta},
 it usually has few consequences.
Further, 
 since our interest is primarily on propagating fields
 in non-trivial media, 
 we leave consideration of interfaces, 
 as well as surface and volume integrals,
 for later work; 
 likewise we do not address the uses of the vector potential $\Vec{A}$
 (or its dual \cite{Hillery-M-1984pra,Datta-1984ejp})
 in this context.

The paper is organized as follows:
 in section \ref{S-maxwell} we introduce Maxwell's equations
 in the forms in which they are used in this paper, 
 in section  \ref{S-continuityeqn} we briefly remark on continuity equations, 
 and
 in section \ref{S-flux} we define our four electromagnetic flux vectors
 and their associated continuity equations.
In section \ref{S-propagation}
 we show how the residual terms modify the underlying 
 propagation of the flux vectors, 
 and we discuss all the results in section \ref{S-discussion}.
Finally, 
 in section \ref{S-conclusion}
 we present our conclusions.

Our presentation is pedagogic in that most of the discussion 
 is at the level of an undergraduate course in electromagnetism. 
The material presented could form extension work
 at the point when Poynting's theorem is introduced. 
Those aspects concerning the propagation of the flux vector
 will be of most interest to specialists.

\section{Maxwell's equations}
\label{S-maxwell}

Maxwell's equations for the electric field $\Vec{E}$
 and magnetic field $\Vec{B}$
 in a medium are
~
\begin{align}
  \grad
 \cdot
  \Vec{E}
&=
  \frac{1}{\epsilon_0}
  \rho_b
 +
  \frac{1}{\epsilon_0}
  \rho_f
 \quad
=
  \frac{1}{\epsilon_0}
  \rho
\\
  \grad
 \cdot
  \Vec{B}
&=
  0
\\
  \grad
 \cross
  \Vec{E}
&=
 -
  \partial_t 
  \Vec{B}
\label{eqn-ME-curlE}
\\
  \grad
 \cross
  \Vec{B}
&=
  \mu_0 
    \Vec{J}_b
 +
  \mu_0 
    \Vec{J}_f
 +
  \mu_0 
  \epsilon_0 
  \partial_t 
  \Vec{E}
\label{eqn-ME-curlB}
,
\end{align}
where $(\rho_b, \Vec{J_b})$ and $(\rho_f, \Vec{J_f})$
 are respectively the bound and free (charge, current) densities.
As an alternative, 
 we can define an electric polarization $\Vec{P}$
 and magnetization $\Vec{M}$, 
 and 
~
\begin{align}
  \Vec{J}_b
&=
  \Vec{J}_P
 +
  \Vec{J}_M
\qquad =
  \partial_t 
  \Vec{P}
 +
  \grad
  \cross
  \Vec{M}
\label{eqn-ME-constit-J}
\\
  \rho_b
&=
 -
  \grad
  \cdot
  \Vec{P}
\label{eqn-ME-divP}
\\
  \Vec{D}
&=
  \epsilon_0
  \Vec{E}
 +
  \Vec{P}
\label{eqn-ME-constit-D}
\\
  \Vec{H}
&=
  \frac{1}{\mu_0}
  \Vec{B}
 -
  \Vec{M}
.
\label{eqn-ME-constit-H}
\end{align}
These allow us to rewrite Maxwell's equations as 
~
\begin{align}
  \grad
 \cdot
  \Vec{D}
&=
  \rho_f
\\
  \grad
 \cdot
  \Vec{B}
&=
  0
\\
  \grad
 \cross
  \Vec{E}
&=
 -
  \mu_0
  \partial_t 
  \left(
    \Vec{H}
   +
    \Vec{M}
  \right)
\label{eqn-MEalt-curlE}
\\
  \grad
 \cross
  \Vec{H}
&=
  \Vec{J}_f
 +
  \partial_t 
  \Vec{D}
\quad
=
  \Vec{J}_f
 +
  \partial_t 
  \left(
    \epsilon_0
    \Vec{E}
   +
    \Vec{P}
  \right)
\label{eqn-MEalt-curlH}
.
\end{align}
We can even rewrite eqn. \eqref{eqn-MEalt-curlE} 
 in the unconventional form
~
\begin{align}
  \grad
 \cross
  \Vec{D}
&=
 -
  \epsilon_0
  \mu_0
  \partial_t
  \left(
    \Vec{H}
   +
    \Vec{M}
  \right)
 +
  \grad
  \cross
  \Vec{P}
\\
&=
 -
  \epsilon_0
  \mu_0
  \partial_t
  \Vec{H}
 -
  \epsilon_0 
  \mu_0
  \Vec{K}_b
,
\label{eqn-MEalt-curlD}
\end{align}
where we have defined 
~
\begin{align}
  \Vec{K}_b
&=
  \Vec{K}_P
 +
  \Vec{K}_M
\qquad =
  -
  \frac{1}{\epsilon_0 \mu_0}
  \grad
  \cross
  \Vec{P}
 +
  \partial_t
  \Vec{M}
,
\label{eqn-ME-monoK}
\\
  \sigma_b
&=
 -
  \grad
  \cdot
  \Vec{M}
\label{eqn-ME-divM}
.
\end{align}
This $\Vec{K}_b$ appears in the same place as a monopole current
 would if such were allowed; 
 $\sigma_b$ is the \emph{bound} magnetic pole density.
Note that $\Vec{K}_b$ and $\sigma_b$
 are merely a way of representing
 the (local) material response; 
 we are not claiming that some process actually 
 generates true magnetic monopoles inside the material
 \cite{RMC}\footnote{Chapter 9, section 3}.
Strictly speaking, 
 this is also true of the bound electric charge and its currents --
 they are a mechanism used solely
 to represent the behaviour of the medium.
Further, 
 and just as for the ficticious bound electric charge density, 
 the ficticious bound monopole density
 necessarily integrates to zero over all space.
Thus the material response could, 
 in principle, 
 be re-represented as magnetic dipoles instead of monopoles.
Note that in using this effective monopole current, 
 we are not going as far as Carpenter \cite{Carpenter-1999ieesmt},
 who posits a complementary universe dominated by magnetic monopoles
 and with no charge in order to clarify some conceptual difficulties.
If we were to include free magnetic monopoles
 and free magnetic monopole currents, 
 then the continuity equations given below would exhibit
 a great deal more symmetry on exchange of electric and magnetic effects.

At this point is is worth noting that the equations above 
 represent the effect of electric polarization in one of two ways:
 either as a current of bound charges
  (i.e. $\Vec{J}_P = \partial_t \Vec{P}$),
 or as a result of bound monopole current loops 
  (i.e. $\Vec{K}_P = \grad \times \Vec{P}$).
They also represent the effect of magnetization similarly:
 either as a current of bound monoples
  (i.e. $\Vec{K}_M = \partial_t \Vec{M}$),
 or as a result of bound electric current loops 
  (i.e. $\Vec{J}_M = \grad \times \Vec{M}$).
Our aversion to free magnetic monopoles, 
 and the widespread acceptance of free electric charges
 may bias many readers toward an electric current picture 
 involving $\Vec{J}_P$ and $\Vec{J}_M$ -- 
 but since these comprise \emph{bound} charges
 which are merely a convenient \emph{fiction}, 
 there is no physical reason not to consider using bound monopoles, 
 or even a mix of the two, 
 if we have sufficient reason.
Indeed, 
 if we (microscopically) model the magnetization as arising
 from some field-induced or environmental distortion  
 of a unit cell or molecule, 
 there is little \emph{a priori} reason not to model 
 the magnetization as induced magnetic dipoles -- 
 we need not take the extra step of assuming 
 the dipoles arise from some induced current loop.

%
\section{Continuity equations}\label{S-continuityeqn}

One of the major uses of flux (Poynting) vectors is in 
 energy continuity equations, 
 where we can examine the balance between flux 
 and local storage of the energy in a medium.
In electromagnetism, 
 the flux vector that is usually chosen
 is the Abraham form of the Poynting vector $\Vec{E} \cross \Vec{H}$, 
 although in some contexts the Minkowski form $\Vec{D} \cross \Vec{B}$
 is chosen.
Some authors
 prefer an $\Vec{E} \cross \Vec{B}$ form for the Poynting vector
 (e.g. the recent \cite{Richter-FH-2008epl}); 
 but only in media with a magnetic response
 does this differ from the Abraham form in anything but scaling.

Our starting point is just a cross-product of two selected fields,
 one electric ($\Vec{E}$ or $\Vec{D}$)
 and one magnetic  ($\Vec{H}$ or $\Vec{B}$).
In concert with Maxwell's equations, 
 such cross-products result in continuity equations 
 of the form
~
\begin{align}
  \grad \cdot \Vec{S}
&=
  \partial_t \mathscr{U}
 +
  \delta
.
\label{eqn-continuity}
\end{align}
Here $\Vec{S}$ is an energy flux
 based on our chosen pair of field vectors.
The energy density $\mathscr{U}$ is a function 
 of the fields used to construct the energy flux vector, 
 and will in general contain all the terms
 that can be expressed as a simple time derivative of some function.
The remaining term $\delta$ is some residual contribution.
One way of avoiding these residual terms
 is to use an energy flux vector of the Umov form, 
 i.e. as an energy density multiplied by a velocity vector, 
 and not the traditional cross-product of fields
 (see e.g. \cite{Crenshaw-A-2006pre}).
In the Umov approach, 
 any terms that are not part of the defined energy density 
 are attributed to the behaviour of the energy flux.

We might now assume that the fields and medium 
 are in some nearly steady state, 
 where the time dependence of $\mathscr{U}$ is either zero,
 or its rapid oscillations average to zero;
 but that the residual terms remain significant.
Here, 
 any change in energy flow $\Vec{S}$
 is balanced by the residual terms $\delta$
 which would most simply be a rate of energy removal or supply.
However, 
 it is perfectly possible for there to be more complicated responses, 
 with a dependence on either time or space; 
 e.g. an in-effect temporary storage of energy
 giving rise to an oscilliatory behaviour.
This is treated more rigorously in section \ref{S-propagation}.

%
\section{Flux vectors}\label{S-flux}

Here we will use the different forms of Maxwell's equations given 
 above to generate four different electromagnetic energy continuity equations.
The way we generate these continuity equations is straightforward: 
 we take our chosen flux vector as defined by a cross product
 of an electric field and a magnetic field
 and take the divergence, 
 using the standard vector identity
~
\begin{align}
  \grad
 \cdot 
  \left(
    \Vec{X}
   \cross
    \Vec{Y}
  \right)
&=
 -
  \Vec{X}
  \cdot
  \left(
    \grad
   \cross
    \Vec{Y}
  \right)
 +
  \Vec{Y}
  \cdot
  \left(
    \grad
   \cross
    \Vec{X}
  \right)
.
\label{eqn-grad-AxB}
\end{align}
Then, 
 by substituting in appropriate Maxwell's equations
 into the RHS to substitute for the curl terms, 
 we generate continuity equations of the general form
 given in eqn. \eqref{eqn-continuity}.
With this style of derivation, 
 there are some subtleties regarding the role
 of total and self-field contributions, 
 as has been discussed by Campoz and Jim\`enez \cite{Campos-J-1992ejp}.

Lastly, 
 if we so wish,
 we might also apply the modifications to our chosen 
 $\Vec{S}_{XY} = \Vec{X} \cross \Vec{Y}$ flux vector that have been applied
 to the (bare) Abraham $\Vec{E} \cross \Vec{H}$ form 
 \cite{Slepian-1942jap,Lai-1981ajp,Peters-1982ajp,Romer-1982ajp,Kobe-1982ajp} 
 --
 e.g. adding terms of zero divergence to $\Vec{S}_{XY}$.

%
\subsection{Abraham $\Vec{E} \cross \Vec{H}$}
\label{S-poynting-ExH}

This Abraham form of electromagnetic flux (Poynting) vector
 is the most widely used of all, 
 consisting of the fields $\Vec{E}$ and $\Vec{H}$.
Inserting $\Vec{E} \cross \Vec{H}$ into the identity 
 eq. \eqref{eqn-grad-AxB}, 
 and using eqns. \eqref{eqn-ME-curlE} and \eqref{eqn-MEalt-curlH}
 gives us a continuity equation, 
 i.e.
~
\begin{align}
  \grad
 \cdot 
  \left(
    \Vec{E}
   \cross
    \Vec{H}
  \right)
&=
 -
  \Vec{E}
  \cdot
  \left(
    \grad
   \cross
    \Vec{H}
  \right)
 +
  \Vec{H}
  \cdot
  \left(
    \grad
   \cross
    \Vec{E}
  \right)
\label{eqn-poynting-ExH-start}
\\
&=
 -
  \Vec{E}
  \cdot
  \partial_t \Vec{D}
 -
  \Vec{E}
  \cdot
  \Vec{J}_f
 -
  \Vec{H}
  \cdot
  \partial_t \Vec{B}
.
\end{align}
This is in itself a widely used expression, 
 but we proceed further to get
~
\begin{align}
  \grad
 \cdot 
  \left(
    \Vec{E}
   \cross
    \Vec{H}
  \right)
&=
 -
  \Vec{E}
  \cdot
  \left(
    \Vec{J}_f
   +
    \epsilon_0
    \partial_t \Vec{E}
   +
    \partial_t \Vec{P}
  \right)
 -
  \Vec{H}
  \cdot
  \mu_0
  \left(
    \partial_t \Vec{H}
   +
    \partial_t \Vec{M}
  \right)
\\
&=
 -
  \frac{1}{2}
  \partial_t 
  \left(
    \epsilon_0
    \Vec{E} \cdot \Vec{E}
   +
    \mu_0
    \Vec{H} \cdot \Vec{H}
  \right)
\nonumber
\\
&\qquad
 -
  \Vec{E}
  \cdot
    \Vec{J}_f
 -
  \Vec{E}
  \cdot
    \partial_t \Vec{P}
 -
  \mu_0
  \Vec{H}
  \cdot
    \partial_t \Vec{M}
\label{eqn-cont-EH-raw}
\\
  \grad
 \cdot
  \Vec{S}_{EH}
&=
 -
  \partial_t 
    \mathscr{U}_{EH}
 -
  \Vec{E}
  \cdot
    \Vec{J}_f
 -
  \Vec{E}
  \cdot
    \partial_t \Vec{P}
 -
  \mu_0
  \Vec{H}
  \cdot
    \partial_t \Vec{M}
,
\label{eqn-cont-EH}
\end{align}
where 
 $\Vec{S}_{EH} = \Vec{E} \cross \Vec{H}$
 and
 $2 \mathscr{U}_{EH} = \epsilon_0 \Vec{E} \cdot \Vec{E}
  + \mu_0 \Vec{H} \cdot \Vec{H} $.

Because we chose the fields $\Vec{E}$ and $\Vec{H}$
 to generate our flux vector, 
 we necessarily find that \emph{both} of these local residual excitations
 depend only on the temporal response of the medium 
 (i.e. they are dispersive).
This means we can replace $\partial_t \Vec{P}$ with 
 the ordinary electric current $\Vec{J}_P$, 
 and $\partial_t \Vec{M}$ with 
 a monopole current contribution $\Vec{K}_M$:
~
\begin{align}
  \grad
 \cdot
  \Vec{S}_{EH}
&=
 -
  \partial_t 
    \mathscr{U}_{EH}
 -
  \Vec{E}
  \cdot
    \Vec{J}_f
 -
  \Vec{E}
  \cdot
    \Vec{J}_P
 -
  \mu_0
  \Vec{H}
  \cdot
    \Vec{K}_M
.
\label{eqn-cont-EH-var}
\end{align}
We cannot somehow introduce $\Vec{J}_M = \grad \cross \Vec{M}$ here
 without making assumptions:
 $\Vec{J}_M$ because that relies on a pre-existing magnetization
 that varies in space.

In optics, 
 researchers work from a starting point
 that is primarily concerned with dispersion; 
 even if that dispersion has usually
 been purely dielectric in origin.
It is natural,
 therefore, 
 for the optics community to prefer the
 $\Vec{E} \cross \Vec{H}$ form, 
 because it treats material polarization and magnetization 
 in a purely temporal manner --
 although we see here that we are then forced
 to represent any magnetic response by bound monopoles.
In the introduction we suggested that the $\Vec{E}$ and $\Vec{H}$ fields
 could be regarded as the bare fields -- 
 and it is this optics inspired temporally-centric view
 in which this is true.
Another nice feature of this choice is that the LHS 
 is a purely spatial derivative 
 (being a divergence $\grad \cdot$), 
 whereas on the RHS both the non-$\Vec{J}_f$ terms 
 (i.e. energy density term and residuals)
 are temporal derivatives.

\subsubsection*{Instantaneous responses}

For the case of a medium with an instantaneous scalar response, 
 we have $\Vec{P} = \chi_P \Vec{E}$, 
 so that $\Vec{E} \cdot \partial_t \Vec{P} 
 \equiv (\partial_t \Vec{E} \cdot \Vec{P})/2$, 
 and so the polarization $\Vec{P}$ 
 can be incorporated into the energy density; 
 the same can be done for the magnetization $\Vec{M}$.
Hence we have 
~
\begin{align}
  \grad
 \cdot 
  \left(
    \Vec{E}
   \cross
    \Vec{H}
  \right)
&=
 -
  \frac{1}{2}
  \partial_t 
  \left[
    \epsilon_0
    \left( 1 + \chi_P \right)
    \Vec{E} \cdot \Vec{E}
   +
    \mu_0
    \left( 1 + \chi_M \right)
    \Vec{H} \cdot \Vec{H}
  \right]
\nonumber
\\
& \qquad
 -
  \Vec{E}
  \cdot
    \Vec{J}_f
\\
&=
 -
  \frac{1}{2}
  \partial_t 
  \left[
    \Vec{E} \cdot \Vec{D}
   +
    \Vec{H} \cdot \Vec{B}
  \right]
 -
  \Vec{E}
  \cdot
    \Vec{J}_f
.
\label{eqn-cont-EH-0}
\end{align}
A consequence of this is that we can move any instantaneous linear
 part of the medium response into the energy density, 
 leaving only the dispersive contributions to remain
 as residual terms.
In practical terms, 
 we can replace $\epsilon_0$ and $\mu_0$
 in eqn. \eqref{eqn-cont-EH} and associated definitions
 by instantaneous-response parameters 
 $\epsilon_i = \epsilon_0 ( 1 + \chi_P )$
 and $\mu_i = \mu_0 ( 1 + \chi_M )$, 
 then only the non-instantaneous or nonlinear responses 
 will remain in $\Vec{P}$ and $\Vec{M}$.

\subsection{Electric current: $\Vec{E} \cross \Vec{B}$}
\label{S-poynting-ExB}

This definition 
 uses the electric field $\Vec{E}$
 and the magnetic field $\Vec{B}$ 
 (see e.g. \cite{Feynman-Lectures-vII,MelroseMcPhedran-EMPDM}),
 and has been recently favoured by some authors for use in complex media
 \cite{Obukhov-H-2003pla,Raabe-W-2005pra,Richter-FH-2008epl,Markel-2008oe}.
However,
 one instance \cite{Raabe-W-2005pra}
 has suffered significant (although initially disputed)
 comment \cite{Pitaevskii-2006pra,Raabe-W-2006pra,Brevik-E-2008pra}
 as to its validity in certain situations.
Leaving aside this dispute,
 we can still use the vector identity above
 to get a continuity equation; 
 this will still be physically valid, 
 even if using it in complex media may mislead the unwary.

To reiterate our earlier point -- 
 rather than attempting to justify a particular choice of flux vector
 with respect to some external criteria, 
 we simply set up the definition(s), 
 and determine what can be done on that basis.
The continuity equation for $\Vec{E} \cross \Vec{B}$
 is based on eqns. \eqref{eqn-ME-curlE} and \eqref{eqn-ME-curlB}, 
 and is
~
\begin{align}
  \grad
 \cdot 
  \left(
    \Vec{E}
   \cross
    \Vec{B}
  \right)
&=
 -
  \Vec{E}
  \cdot
  \left(
    \grad
   \cross
    \Vec{B}
  \right)
 +
  \Vec{B}
  \cdot
  \left(
    \grad
   \cross
    \Vec{E}
  \right)
\label{eqn-poynting-ExB-start}
\\
&=
 -
  \Vec{E}
  \cdot
  \left(
    \mu_0
    \Vec{J}
   +
    \mu_0
    \epsilon_0
    \partial_t \Vec{E}
  \right)
 -
  \Vec{B}
  \cdot
    \partial_t \Vec{B}
\\
&=
 -
  \frac{\mu_0}{2}
  \partial_t 
  \left(
    \epsilon_0
    \Vec{E} \cdot \Vec{E}
   +
    \frac{1}{\mu_0}
    \Vec{B} \cdot \Vec{B}
  \right)
 -
  \mu_0
  \Vec{E}
  \cdot
    \Vec{J}
\label{eqn-cont-EB-raw}
\\
  \grad
 \cdot
  \Vec{S}_{EB}
&=
 -
  \partial_t 
    \mathscr{U}_{EB}
 -
  \Vec{E}
  \cdot
    \Vec{J}
,
\label{eqn-cont-EB}
\end{align}
where of course $\Vec{J} = \Vec{J}_f + 
 \partial_t \Vec{P} + \grad \cross \Vec{M}$; 
 this is just the total current density; 
 also
 $\mu_0 \Vec{S}_{EB} = \Vec{E} \cross \Vec{B}$
 and
 $2 \mathscr{U}_{EB} = \epsilon_0 \Vec{E} \cdot \Vec{E}
  + \Vec{B} \cdot \Vec{B}/\mu_0 $.

The simple representation of the residual terms above
 (i.e. as $\Vec{E} \cdot \Vec{J}$)
 is presumably the underlying reason why $\Vec{E} \cross \Vec{B}$
 is preferred by many authors: 
 the material response is represented in terms of an electric current, 
 and $\Vec{E} \cdot \Vec{J}$ can be interpreted simply 
 as the work done on charges.
However, 
 from a more general point of view, 
 $\Vec{E} \cross \Vec{B}$ is not the ``correct'' form 
 of the Poynting vector, 
 it just is one of many,
 each of which may (or may not) be more convenient 
 in a particular situation.
Note that here only the polarization residual excitation
 depends on the temporal response (i.e. dispersion); 
 the magnetic part is purely spatial:
 both forms combine to generate a bound electric current.
This is obviously an advantage to those who
 prefer to think only in terms of electric charges,
 whether real or ficticious; 
 certainly $\Vec{E} \cross \Vec{B}$ is the obvious choice
 for any \emph{microscopic} model involving real charges
 (and no real monoples).
However, 
 the time response (dispersion) of the magnetization is no longer explicit, 
 as it was for $\Vec{E} \cross \Vec{H}$,
 but has become hidden inside the bound magnetization current $\Vec{J}_M$.

In the previous subsection,
 we suggested that the $\Vec{E}$ and $\Vec{H}$ fields
 could be regarded as the bare fields.
This is a point of view often taken in optics,
 and is consistent with the $\Vec{E} \cross \Vec{H}$ choice
 of Poynting vector.
However, 
 choosing $\Vec{E} \cross \Vec{B}$
 gives us an alternative electric charge (or current) centric view, 
 in which case it is the $\Vec{E}$ and $\Vec{B}$ fields
 which are regarded as the bare fields, 
 indeed they are already widely regarded
 as the fundamental electromagnetic fields
 \cite{Feynman-Lectures-vII}\footnote{Chapter 27, section 3}
 and \cite{MelroseMcPhedran-EMPDM}\footnote{Chapter 1, section 1}.

\subsubsection*{Instantaneous and pointlike responses}

In the case of instantaneous polarization response
 and pointlike magnetic response, 
 we find that eqn. \eqref{eqn-poynting-ExB-start}
 can be directly reduced 
 to the same result as in eqn. \eqref{eqn-cont-EH-0},
 but scaled by $\mu_0$, 
 since $\Vec{H}$ has been replaced by $\Vec{B}$ in the flux vector.
A consequence of this is that we can move such
 components of the medium response into the energy density, 
 leaving only the dispersive polarization
 and spatial magnetic parts as residual terms.

\subsection{Magnetic current: $\Vec{D} \cross \Vec{H}$}
\label{S-poynting-DxH}

This alternate,
 and little (or never) used definition 
 comprises the magnetic induction field $\Vec{H}$
 and the displacement field $\Vec{D}$
 as its basic components.
Note that the historical review
  of Buchwald \cite{Buchwald-FMTM} makes some discouraging remarks 
  as regards choosing $\Vec{D}$ and $\Vec{H}$ as fundamental fields 
  at the end of chapter 2 (p.18), 
  but makes no reference to a flux generated from them.
Alternatively, 
 the projection approach taken to pulse propagation 
 by Kolesik et al. \cite{Kolesik-MM-2002prl,Kolesik-M-2004pre}
 relied on the $\Vec{D}$ and $\Vec{H}$ fields, 
 although they did not consider media with a magnetic response.
The $\Vec{D} \cross \Vec{H}$ continuity equation
 is based on eqns. \eqref{eqn-MEalt-curlH}, \eqref{eqn-MEalt-curlD}, 
 and is
~
\begin{align}
  \grad
 \cdot 
  \left(
    \Vec{D}
   \cross
    \Vec{H}
  \right)
&=
 -
  \Vec{D}
  \cdot
  \left(
    \grad
   \cross
    \Vec{H}
  \right)
 +
  \Vec{H}
  \cdot
  \left(
    \grad
   \cross
    \Vec{D}
  \right)
\label{eqn-poynting-DxH-start}
\\
&=
 -
  \Vec{D}
  \cdot
  \left(
    \Vec{J}_f
   +
    \partial_t \Vec{D}
  \right)
\nonumber
\\
&\qquad
 -
  \Vec{H}
  \cdot
  \left(
    \epsilon_0 \mu_0
    \partial_t \Vec{H}
   +
    \epsilon_0 \mu_0
    \partial_t \Vec{M}
   -
    \grad \cross \Vec{P}
  \right)
\\
&=
 -
  \frac{\epsilon_0}{2}
  \partial_t 
  \left(
    \frac{1}{\epsilon_0}
    \Vec{D} \cdot \Vec{D}
   +
    \mu_0
    \Vec{H} \cdot \Vec{H}
  \right)
\nonumber
\\
&\qquad
 -
  \Vec{D}
  \cdot
    \Vec{J}_f
 -
  \epsilon_0
  \mu_0
  \Vec{H}
  \cdot
  \partial_t \Vec{M}
 +
  \Vec{H}
  \cdot 
    \grad \cross \Vec{P}
\label{eqn-cont-DH-raw}
\\
  \grad
 \cdot
  \Vec{S}_{DH}
&=
 -
  \partial_t 
    \mathscr{U}_{DH}
 -
  \frac{1}{\epsilon_0}
  \Vec{D}
  \cdot
    \Vec{J}_f
 -
  \mu_0
  \Vec{H}
  \cdot
  \partial_t \Vec{M}
 +
  \frac{1}{\epsilon_0}
  \Vec{H}
  \cdot
  \grad \cross \Vec{P}
,
\label{eqn-cont-DH}
\end{align}
where 
 $\epsilon_0 \Vec{S}_{DH} = \Vec{D} \cross \Vec{H}$
 and
 $2 \mathscr{U}_{DH} = \Vec{D} \cdot \Vec{D} / \epsilon_0 
  + \mu_0 \Vec{H} \cdot \Vec{H}$.
Note that only the magnetization residual excitation
 depends on the temporal response (i.e. dispersion); 
 the polarization part is purely spatial; 
 as a result the material response can be encoded solely
 by means of the bound monopole current $\Vec{K}_b$
 defined in eqn. \eqref{eqn-ME-monoK}:
~
\begin{align}
  \grad
 \cdot
  \Vec{S}_{DH}
&=
 -
  \partial_t 
    \mathscr{U}_{DH}
 -
  \frac{1}{\epsilon_0}
  \Vec{D}
  \cdot
    \Vec{J}_f
 -
  \mu_0
  \Vec{H}
  \cdot
  \Vec{K}_b
\label{eqn-cont-DH-var}
,
\end{align}
although the free current part still mimics
 the $\Vec{E} \cdot \Vec{J}$ form.

This $\Vec{D} \cross \Vec{H}$ form, 
 therefore, 
 is the natural complement
 to the electric current based $\Vec{E} \times \Vec{B}$ form, 
 because the material response in both
 is completely encoded by means of a current:
 but one is a magnetic monople current $\Vec{K}_b$, 
 and one an ordinary electric current $\Vec{J}_b$.
It is probably unsurprising, 
 therefore, 
 that $\Vec{D} \cross \Vec{H}$ has been neglected, 
 because it treats material polarization and magnetization 
 purely as monopole currents --
 even though these are fictitious monopoles, 
 bound in dipole pairs, 
 introduced solely to represent (model) the material response.
Indeed, 
 the description here is a purely continuum one, 
 and the use of a monopole current is not
 contingent on the existence of fundamental particles 
 carrying magnetic charge.

The $\Vec{E} \times \Vec{H}$ Poyting vector
 gave us a time-centric viewpoint, 
 and was compatible with regarding $\Vec{E}$ and $\Vec{H}$
 as the bare fields, 
 similarly the electric-current centred viewpoint of $\Vec{E} \times \Vec{B}$
 led us to insist that $\Vec{E}$ and $\Vec{B}$
 are the bare fields.
Here we have seen that a monopole-current centred viewpoint 
 might encourage the idea that $\Vec{D}$ and $\Vec{H}$
 are the bare electromagnetic fields!

\subsubsection*{Pointlike and instantaneous responses}

In the case of pointlike polarization response
 and instantaneous magnetic response, 
 we find that eqn. \eqref{eqn-poynting-DxH-start}
 can be directly reduced 
 to the same result as in eqn. \eqref{eqn-cont-EH-0},
 but scaled by $\epsilon_0$.
A consequence of this is that we can move such
 components of the medium response into the energy density, 
 leaving only the dispersive magnetic
 and spatial polarization parts as residual terms.

\subsection{Minkowski $\Vec{D} \cross \Vec{B}$}
\label{S-poynting-DxB}

This uses the usual definition 
 of the Minkowski Poynting vector, 
 which contains the 
 displacement and magnetic fields $\Vec{D}$ and $\Vec{B}$.
Its continuity equation
 is based on eqns. \eqref{eqn-ME-curlB} and \eqref{eqn-MEalt-curlD}, 
 along with eqns. \eqref{eqn-ME-constit-J} and \eqref{eqn-ME-constit-D}, 
 and is
~
\begin{align}
  \grad
 \cdot 
  \left(
    \Vec{D}
   \cross
    \Vec{B}
  \right)
&=
 -
  \Vec{D}
  \cdot
  \left(
    \grad
   \cross
    \Vec{B}
  \right)
 +
  \Vec{B}
  \cdot
  \left(
    \grad
   \cross
    \Vec{D}
  \right)
\label{eqn-poynting-DxB-start}
\\
&=
 -
  \Vec{D}
  \cdot
  \mu_0
  \left(
    \Vec{J}_f + \grad \cross \Vec{M} + \partial_t \Vec{D}
  \right)
\nonumber
\\
&\qquad
 +
  \Vec{B}
  \cdot
  \left(
   -
    \epsilon_0
    \partial_t \Vec{B}
   +
    \grad
    \cross
    \Vec{P}
  \right)
\\
&=
 -
  \frac{\mu_0\epsilon_0}{2}
  \partial_t 
  \left(
    \frac{1}{\epsilon_0}
    \Vec{D} \cdot \Vec{D}
   +
    \frac{1}{\mu_0}
    \Vec{B} \cdot \Vec{B}
  \right)
\nonumber
\\
&\qquad
 -
  \mu_0
  \Vec{D} 
  \cdot 
  \left( \Vec{J}_f + \grad \cross \Vec{M} \right)
 +
  \Vec{B}
  \cdot
    \grad
    \cross
    \Vec{P}
\label{eqn-cont-DB-raw}
\\
  \grad
 \cdot
  \Vec{S}_{DB}
&=
 -
  \partial_t 
    \mathscr{U}_{DB}
 -
  \frac{1}{\epsilon_0}
  \Vec{D} 
  \cdot 
  \left( \Vec{J}_f + \grad \cross \Vec{M} \right)
 +
  \frac{1}{\epsilon_0\mu_0}
  \Vec{B}
  \cdot
    \grad
    \cross
    \Vec{P}
,
\label{eqn-cont-DB}
\end{align}
where 
 $\epsilon_0 \mu_0 \Vec{S}_{DB} = \Vec{D} \cross \Vec{B}$
 and
 $2 \mathscr{U}_{DB} = \Vec{D} \cdot \Vec{D} / \epsilon_0 
  + \Vec{B} \cdot \Vec{B}/\mu_0 $.

Note that neither residual excitation
 depends on the temporal response (i.e. dispersion); 
 the polarization and magnetization parts are purely spatial.
This means that we might replace $\grad \cross \Vec{M}$ with 
 the electric current $\Vec{J}_M$, 
 and $\grad \cross \Vec{P}$ with 
 the monopole current $\Vec{K}_P/\epsilon_0\mu_0$:
~
\begin{align}
  \grad
 \cdot
  \Vec{S}_{DB}
&=
 -
  \partial_t 
    \mathscr{U}_{DB}
 -
  \frac{1}{\epsilon_0}
  \Vec{D} 
  \cdot 
  \left( \Vec{J}_f + \Vec{J}_M \right)
 -
  \Vec{B}
  \cdot
  \Vec{K}_P
\label{eqn-cont-DB-var}
.
\end{align}

This form, 
 therefore, 
 treats material polarization and magnetization 
 in a purely spatial manner; 
 as such it promotes a picture wherein 
 it is the $\Vec{D}$ and $\Vec{B}$ fields
 which look like the bare fields.

\subsubsection*{Pointlike responses}

In the case of pointlike magnetic and polarization responses, 
 we find that eqn. \eqref{eqn-poynting-DxB-start}
 can be directly reduced 
 to the same result as in eqn. \eqref{eqn-cont-EH-0},
 but scaled by $\epsilon_0 \mu_0$, 
 since $\Vec{E}$ and $\Vec{H}$ have been replaced by $\Vec{D}$ and $\Vec{B}$.
A consequence of this is that we can move such
 components of the medium response into the energy density, 
 leaving simplified residual terms.

%
\section{Propagation of flux}
\label{S-propagation}

We can now show how these residual terms
 affect a propagating wave, 
 by deriving propagation equations for the Poynting vectors themselves.
To do this we use the concept of directional electromagnetic fields
 \cite{Kinsler-RN-2005pra,Kinsler-2006arXiv-fleck,Mizuta-NOY-2005pra}, 
 and as a result do not need to 
 resort to restrictive approximations, 
 such as e.g. assuming plane wave or harmonic fields.
Notably, 
 we will not need to resort to ad-hoc time averaging
 of fast-oscillating terms, 
 as done in this kind of context by e.g. Markel and others
 \cite{Markel-2008oe,Favaro-KM-2009oe};
 neither do we need to introduce pulse envelopes, 
 co-moving frames, 
 or make smoothness assumptions 
 \cite{Boyd-NLO,Brabec-K-1997prl,Geissler-TSSKB-1999prl,Kinsler-N-2003pra}.

First we note that 
 each of our electromagnetic continuity 
 eqns. \eqref{eqn-cont-EH-raw},
 \eqref{eqn-cont-EB-raw},
 \eqref{eqn-cont-DH-raw},
 \eqref{eqn-cont-DB-raw},
 has the general form
~
\begin{align}
 -
  c
  \grad
 \cdot
  \left(
   \Vec{X} \cross \Vec{Y}
  \right)
&=
  \frac{1}{2}
  \partial_t
  \left[
    \Vec{X} \cdot \Vec{X}
   +
    \Vec{Y} \cdot \Vec{Y}
  \right]
 +
  \Vec{R}
.
\end{align}
Now assume transverse fields propagating in the direction
 given by a unit vector $\Vec{u}$, 
 so that $\Vec{X} \cdot \Vec{u} = \Vec{Y} \cdot \Vec{u} = 0$.
This means we can construct directional fields
 \cite{Kinsler-RN-2005pra,Kinsler-2006arXiv-fleck} by defining
~
\begin{align}
  \Vec{G}^\pm
&=
  \Vec{X}
 \mp
   \Vec{u} \cross \Vec{Y}
;
\end{align}
where e.g. 
 for plane polarized fields we might use 
 $X_x = \sqrt{\epsilon_0} E_x = D_x / \sqrt{\epsilon_0}$ 
 and $Y_y = \sqrt{\mu_0} H_y = B_y / \sqrt{\mu_0}$; 
 however note that we need not be restricted to 
 scaling by the vacuum values of $\epsilon_0$ and $\mu_0$.
Thus since
 $\Vec{X} \cross \Vec{Y} 
    = \Vec{u} ( |\Vec{G}^{+}|^2 - |\Vec{G}^{-}|^2 ) / 4$, 
 we can write
~
\begin{align}
 -
  c
  \grad
  \cdot
  \Vec{u}
  \left[
    \left| \Vec{G}^{+}\right|^2 
   -
    \left| \Vec{G}^{-} \right|^2 
  \right]
&=
  \frac{1}{4}
  \partial_t
  \left[
    \left| \Vec{G}^{+}\right|^2 
   +
    \left| \Vec{G}^{-} \right|^2 
  \right]
 +
  \Vec{R}
,
\end{align}
which is easily rearranged to
~
\begin{align}
  \left[
    \partial_t
   -
    c
    \grad \cdot \Vec{u}
  \right]
    \left| \Vec{G}^{+}\right|^2 
 +
  \left[
    \partial_t
   +
    c
    \grad \cdot \Vec{u}
  \right]
    \left| \Vec{G}^{-} \right|^2 
&= 
 -
  4
  \Vec{R}
.
\label{eqn-G2-bi}
\end{align}
This contains two counter-propagating components, 
 each evolved by its own wave operator $\partial_t \mp c \grad \cdot \Vec{u}$, 
 with the residual terms remaining on the RHS.
If we are only interested in unidirectional propagation
 we can set $\Vec{G}^-=0$, 
 and with $\Vec{u} \parallel \hat{z}$, 
 we get
~
\begin{align}
  \left[
    \partial_t
   -
    c
    \partial_z
  \right]
    \left| \Vec{G}^{+}\right|^2 
&= 
 -
  4
  \Vec{R}
.
\label{eqn-G2-uni}
\end{align}
This is a simple first order wave equation for the intensity, 
 and we can easily see that loss-like residual terms $\Vec{R}$
 will cause that intensity to diminish; 
 similarly those dependent on the past 
 can cause dispersion.
However, 
 in the same way as with standard directional fields \cite{Kinsler-RN-2005pra}
 or factorization approaches 
 \cite{Kinsler-2007-envel,Genty-KKD-2007oe}, 
 we require that the residual terms only have a small effect 
 over the scale of one wavelength in order for $\Vec{G}^-$ to stay negligible
 \cite{Kinsler-2007josab,Kinsler-2008-fbdiff,Kinsler-2010pra-fchhg}.

Note that a considerable amount of detail is hidden
 in the total residual term $\Vec{R}$; 
 the various residual components are summarized in table \ref{tab-summary}
 for each combination of field vectors.
If we wanted to solve eqn. \eqref{eqn-G2-uni}
 (or perhaps even \eqref{eqn-G2-bi}),
 we would need to rewrite those residual components
 in terms of $\Vec{G}^{\pm}$; 
 the approximation $\Vec{G}^-=0$ helps in this respect 
 since it allows the field choice $\Vec{X}$ 
 to be expressed in terms of $\Vec{Y}$.
In any case, 
 we can see that the components of $\Vec{R}$
 will act as source terms that drive and modify
 the otherwise simple linear wave propagation.

\begin{widetext} 

\begin{table}[h]
 \caption{Summary of flux vectors and their
 corresponding energy densities, 
 along with the residual terms 
 and their corresponding bound currents.}
 \begin{tabular}{lcccc}
 \hline 
   Flux \qquad \qquad & \qquad Energy density \qquad & \qquad Temporal \qquad & \qquad Spatial \qquad & \qquad Currents \qquad\\
 \hline
   $\Vec{E} \cross \Vec{H}$ &
   \qquad$\epsilon_0 \Vec{E} \cdot \Vec{E} + \mu_0 \Vec{H} \cdot \Vec{H}$\qquad &
   \qquad$\Vec{E} \cdot \partial_t \Vec{P} + \Vec{H} \cdot \partial_t \Vec{M}$\qquad &
   0 &
   \qquad$\Vec{J}_P, \Vec{K}_M$\qquad
\\
   $\Vec{E} \cross \Vec{B}$ &
   \qquad$\epsilon_0 \Vec{E} \cdot \Vec{E} + \mu_0^{-1}\Vec{B} \cdot \Vec{B}$\qquad &
   \qquad$\Vec{E} \cdot \partial_t \Vec{P}$\qquad &
   \qquad$+\Vec{E} \cdot \grad \cross\Vec{M}$\qquad &
   \qquad$\Vec{J}_b$\qquad
\\
   $\Vec{D} \cross \Vec{H}$ &
   \qquad$\epsilon_0^{-1} \Vec{D} \cdot \Vec{D} + \mu_0 \Vec{H} \cdot \Vec{H}$\qquad &
   \qquad$\Vec{H} \cdot \partial_t \Vec{M}$\qquad &
   \qquad$-\Vec{H} \cdot \grad \cross \Vec{P}$\qquad &
   \qquad$\Vec{K}_b$\qquad
\\
   $\Vec{D} \cross \Vec{B}$ &
   \qquad$\epsilon_0^{-1} \Vec{D} \cdot \Vec{D} + \mu_0^{-1} \Vec{B} \cdot \Vec{B}$\qquad &
   0 &
   \qquad$- \Vec{B} \cdot \grad \cross \Vec{P}
    + \Vec{D} \cdot \grad \cross \Vec{M}$\qquad &
   \qquad$\Vec{K}_P, \Vec{J}_M$\qquad
\\
\hline
\end{tabular}
 \label{tab-summary}
\end{table} 

\end{widetext}

%
\section{Discussion}
\label{S-discussion}

Here we compare and contrast the residual terms, 
 which neither appear in the form of an energy flux nor an energy density; 
 they are summarized in table \ref{tab-summary}.
The Abraham $\Vec{E} \cross \Vec{H}$ form 
 has residuals that are purely dispersive, 
 i.e. depend on the temporal derivatives of the medium responses
 via $\Vec{P}$ and $\Vec{M}$.
In contrast, 
 the Minkowski $\Vec{D} \cross \Vec{B}$ form 
 has residuals that are purely spatial 
 i.e. depend on the curl of the medium responses.
The lesser used forms $\Vec{E} \cross \Vec{B}$ and $\Vec{D} \cross \Vec{H}$
 have mixed residuals -- 
 the $\Vec{E} \cross \Vec{B}$ residuals conveniently match the form 
 of the usual definition of the total current; 
 whereas the $\Vec{D} \cross \Vec{H}$ residuals 
 match the less conventional picture of a material response 
 described by bound monopoles.

Let us now consider the effects of these residual terms
 in all four cases, 
 where we assume that $\Vec{J}_f=0$.


For $\Vec{E} \cross \Vec{H}$, 
 the continuity eqn. \eqref{eqn-cont-EH-raw}
 shows that the flux vs energy density balance 
 is modified by two residual terms, 
 i.e. the local dielectric excitation 
 $\Vec{E} \cdot \partial_t \Vec{P}$, 
 and local magnetic excitation 
 $\Vec{H} \cdot \partial_t \Vec{M}$.
Both of these are temporal, 
 so if we choose to propagate forward in space, 
 which is a common choice when considering propagation in optics
 (see e.g. \cite{Boyd-NLO,Kinsler-2007-envel}), 
 during each step forward in space, 
 the fields $\Vec{E}(t), \Vec{H}(t)$ will be modified 
 according to their time derivatives.


For $\Vec{E} \cross \Vec{B}$,
 the continuity eqn. \eqref{eqn-cont-EB-raw}
 shows that the flux vs energy density balance 
 is modified by two residual terms, 
 one temporal $\Vec{E} \cdot \partial_t \Vec{P}$,
 and one spatial $\Vec{E} \cdot \grad \cross \Vec{M}$; 
 in concert they represent the local electrical work 
 done on the charges comprising
 the current $\Vec{J}$.
Unlike in the optics $\Vec{E} \cross \Vec{H}$ picture, 
 here a propagation step forward in space 
 would be complicated by the spatial term; 
 likewise the alternate choice of a time propagation step
 would be complicated by the temporal term.


For $\Vec{D} \cross \Vec{H}$,
 the continuity eqn. \eqref{eqn-cont-DH-raw}
 shows that the flux vs energy density balance 
 is modified by two residual terms, 
 one spatial $\Vec{H} \cdot \grad \cross \Vec{P}$,
 and one temporal $-\Vec{H} \cdot \partial_t \Vec{M}$;
 in concert they represent the local magnetic work 
 done on the monople current $\Vec{K}$.
Just as for the $\Vec{E} \cross \Vec{B}$ picture, 
 this leads to complicated methodology
 for propagation of the fields.


For $\Vec{D} \cross \Vec{B}$, 
 the continuity eqn. \eqref{eqn-cont-DB-raw}
 shows that the flux vs energy density balance 
 is modified by two residual terms, 
 both spatial:
 $-\Vec{B} \cdot \grad \cross \Vec{P}$,
 and $\Vec{D} \cdot \grad \cross \Vec{M}$.
Just as for the $\Vec{E} \cross \Vec{H}$ picture, 
 both residuals are of the same type 
 (albiet spatial, not temporal), 
 so that making the ``causal'' choice \cite{Kinsler-2011ejp}
 to propagate in time
 gives us the simple case where
 the fields $\Vec{D}(\Vec{r}), \Vec{B}(\Vec{r})$ will be modified 
 according to their spatial derivatives \cite{Kinsler-2015arxiv-d2owe}.
Despite this, 
 propagation treatments that evolve the fields forward in time
 (e.g. FDTD \cite{Yee-1966tap}, or \cite{Scalora-C-1994oc,Kolesik-M-2004pre})
 generally tend to persist with the use of $\Vec{E}$ and $\Vec{H}$.

%
\section{Conclusion}\label{S-conclusion}

We have summarized four distinct electromagnetic continuity equations, 
 each being derived from
 (and being consistent with) 
 Maxwell's equations
 and the standard constitutive relations.
Each handles the electric or magnetic response of the medium
 in a slightly different way, 
 so the most appropriate form needs to be chosen according
 to the system under study.

Not only have we presented the different interpretations 
 motivated by each of the Abraham ($\Vec{E} \cross \Vec{H}$),
 Minkowski ($\Vec{D} \cross \Vec{B}$), 
 and standard electromagnetic (``electric current'')
 $\Vec{E} \cross \Vec{B}$ forms, 
 but we also consider the alternative (``magnetic current'')
 $\Vec{D} \cross \Vec{H}$ form as well. 
Finally, 
 we showed how a directional fields approach \cite{Kinsler-RN-2005pra}
 could be used 
 to generate a propagation equation for each flux vector.
This propagation equation makes a relatively small number of assumptions,
 and so not only enables further interpretation 
 of each flux and its interaction with the medium, 
 but has potential applications in its own right --
 even in materials with a complex response \cite{FocusIssue-2003oe-nrm},
 and for wideband or ultrafast optical pulses \cite{Brabec-K-2000rmp}.


\acknowledgments

The authors acknowledge financial support from the
 Engineering and Physical Sciences Research Council
 (EP/E031463/1).


\bibliography{/home/physics/_work/bibtex.bib}

\begin{thebibliography}{49}
\expandafter\ifx\csname natexlab\endcsname\relax\def\natexlab#1{#1}\fi
\expandafter\ifx\csname bibnamefont\endcsname\relax
  \def\bibnamefont#1{#1}\fi
\expandafter\ifx\csname bibfnamefont\endcsname\relax
  \def\bibfnamefont#1{#1}\fi
\expandafter\ifx\csname citenamefont\endcsname\relax
  \def\citenamefont#1{#1}\fi
\expandafter\ifx\csname url\endcsname\relax
  \def\url#1{\texttt{#1}}\fi
\expandafter\ifx\csname urlprefix\endcsname\relax\def\urlprefix{URL }\fi
\providecommand{\bibinfo}[2]{#2}
\providecommand{\eprint}[2][]{\url{#2}}

\bibitem[{\citenamefont{Abraham}(1909)}]{Abraham-1909rcmp}
\bibinfo{author}{\bibfnamefont{M.}~\bibnamefont{Abraham}},
  \\ \bibinfo{journal}{Rend. Circ. Mat. Palermo} \textbf{\bibinfo{volume}{28}},
  \bibinfo{pages}{1} (\bibinfo{year}{1909}).

\bibitem[{\citenamefont{Abraham}(1910)}]{Abraham-1910rcmp}
\bibinfo{author}{\bibfnamefont{M.}~\bibnamefont{Abraham}},
  \\ \bibinfo{journal}{Rend. Circ. Mat. Palermo} \textbf{\bibinfo{volume}{30}},
  \bibinfo{pages}{33} (\bibinfo{year}{1910}).

\bibitem[{\citenamefont{Minkowski}(1910)}]{Minkowski-1908}
\bibinfo{author}{\bibfnamefont{H.}~\bibnamefont{Minkowski}},
  \\ \bibinfo{journal}{Math. Ann.} \textbf{\bibinfo{volume}{68}},
  \bibinfo{pages}{472} (\bibinfo{year}{1910}), \\ \bibinfo{note}{reprint of
  Minkowski, H., 1908, Nachr. Ges. Wiss. Goettingen, Math.- Phys. Kl. 1908,
  53.}

\bibitem[{\citenamefont{Pfeifer et~al.}(2007)\citenamefont{Pfeifer, Nieminen,
  Heckenberg, and Rubinsztein-Dunlop}}]{Pfeifer-NHR-2007rmp}
\bibinfo{author}{\bibfnamefont{R.~N.~C.} \bibnamefont{Pfeifer}},
  \bibinfo{author}{\bibfnamefont{T.~A.} \bibnamefont{Nieminen}},
  \bibinfo{author}{\bibfnamefont{N.~R.} \bibnamefont{Heckenberg}},
  \bibnamefont{and}
  \bibinfo{author}{\bibfnamefont{H.}~\bibnamefont{Rubinsztein-Dunlop}},
  \\ \bibinfo{journal}{Rev. Mod. Phys.} \textbf{\bibinfo{volume}{79}},
  \bibinfo{pages}{1197} (\bibinfo{year}{2007}), \\ \XARXIV{0710.0461},
  \\ \XDOI{10.1103/RevModPhys.79.1197}.

\bibitem[{\citenamefont{Poynting}(1884)}]{Poynting-1884rspl}
\bibinfo{author}{\bibfnamefont{J.~H.} \bibnamefont{Poynting}},
  \\ \bibinfo{journal}{Proceedings of the Royal Society of London}
  \textbf{\bibinfo{volume}{38}}, \bibinfo{pages}{168} (\bibinfo{year}{1884}),
  \\ \XDOI{10.1098/rspl.1884.0080}.

\bibitem[{\citenamefont{Landau and Lifshitz}(1984)}]{LandauLifshitz}
\bibinfo{author}{\bibfnamefont{L.~D.} \bibnamefont{Landau}} \bibnamefont{and}
  \bibinfo{author}{\bibfnamefont{E.~M.} \bibnamefont{Lifshitz}},
  \emph{\bibinfo{title}{Electrodynamics of Continuous Media}}
  (\bibinfo{publisher}{Pergamon}, \bibinfo{address}{Oxford and New York},
  \bibinfo{year}{1984}).

\bibitem[{\citenamefont{Slepian}(1942)}]{Slepian-1942jap}
\bibinfo{author}{\bibfnamefont{J.}~\bibnamefont{Slepian}}, \\ \bibinfo{journal}{J.
  Appl. Phys.} \textbf{\bibinfo{volume}{13}}, \bibinfo{pages}{512}
  (\bibinfo{year}{1942}), \\ \XDOI{10.1063/1.1714903}.

\bibitem[{\citenamefont{Lai}(1981)}]{Lai-1981ajp}
\bibinfo{author}{\bibfnamefont{C.~S.} \bibnamefont{Lai}}, \\ \bibinfo{journal}{Am.
  J. Phys.} \textbf{\bibinfo{volume}{49}}, \bibinfo{pages}{841}
  (\bibinfo{year}{1981}), \\ \XDOI{10.1119/1.12719}.

\bibitem[{\citenamefont{Peters}(1982)}]{Peters-1982ajp}
\bibinfo{author}{\bibfnamefont{P.~C.} \bibnamefont{Peters}},
  \\ \bibinfo{journal}{Am. J. Phys.} \textbf{\bibinfo{volume}{50}},
  \bibinfo{pages}{1165} (\bibinfo{year}{1982}),
  \\ \XDOI{10.1119/1.13024}.

\bibitem[{\citenamefont{Romer}(1982)}]{Romer-1982ajp}
\bibinfo{author}{\bibfnamefont{R.~H.} \bibnamefont{Romer}},
  \\ \bibinfo{journal}{Am. J. Phys.} \textbf{\bibinfo{volume}{50}},
  \bibinfo{pages}{1166} (\bibinfo{year}{1982}),
  \\ \XDOI{10.1119/1.12903}.

\bibitem[{\citenamefont{Kobe}(1982)}]{Kobe-1982ajp}
\bibinfo{author}{\bibfnamefont{D.~H.} \bibnamefont{Kobe}},
  \\ \bibinfo{journal}{Am. J. Phys.} \textbf{\bibinfo{volume}{50}},
  \bibinfo{pages}{1162} (\bibinfo{year}{1982}),
  \\ \XDOI{10.1119/1.12901}.

\bibitem[{\citenamefont{Feynman et~al.}(1964)\citenamefont{Feynman, Leighton,
  and Sands}}]{Feynman-Lectures-vII}
\bibinfo{author}{\bibfnamefont{R.~P.} \bibnamefont{Feynman}},
  \bibinfo{author}{\bibfnamefont{R.~B.} \bibnamefont{Leighton}},
  \bibnamefont{and} \bibinfo{author}{\bibfnamefont{M.}~\bibnamefont{Sands}},
  \emph{\bibinfo{title}{The Feynman lectures on physics}},
  vol.~\bibinfo{volume}{2} (\bibinfo{publisher}{Addison-Wesley},
  \bibinfo{year}{1964}), \\ \bibinfo{note}{14th printing}.

\bibitem[{\citenamefont{Melrose and McPhedran}(1991)}]{MelroseMcPhedran-EMPDM}
\bibinfo{author}{\bibfnamefont{D.~B.} \bibnamefont{Melrose}} \bibnamefont{and}
  \bibinfo{author}{\bibfnamefont{R.~C.} \bibnamefont{McPhedran}},
  \emph{\bibinfo{title}{Electromagnetic processes in dispersive media}}
  (\bibinfo{publisher}{Cambridge University Press},
  \bibinfo{address}{Cambridge}, \bibinfo{year}{1991}), ISBN
  \bibinfo{isbn}{0-521-41025-8}.

\bibitem[{\citenamefont{Reitz et~al.}(1980)\citenamefont{Reitz, Milford, and
  Christy}}]{RMC}
\bibinfo{author}{\bibfnamefont{J.~R.} \bibnamefont{Reitz}},
  \bibinfo{author}{\bibfnamefont{F.~J.} \bibnamefont{Milford}},
  \bibnamefont{and} \bibinfo{author}{\bibfnamefont{R.~W.}
  \bibnamefont{Christy}}, \emph{\bibinfo{title}{Foundations of electromagnetic
  theory}} (\bibinfo{publisher}{Addison-Wesley}, \bibinfo{year}{1980}),
  \bibinfo{edition}{3rd} ed.

\bibitem[{\citenamefont{Richter et~al.}(2008)\citenamefont{Richter, Florian,
  and Henneberger}}]{Richter-FH-2008epl}
\bibinfo{author}{\bibfnamefont{F.}~\bibnamefont{Richter}},
  \bibinfo{author}{\bibfnamefont{M.}~\bibnamefont{Florian}}, \bibnamefont{and}
  \bibinfo{author}{\bibfnamefont{K.}~\bibnamefont{Henneberger}},
  \\ \bibinfo{journal}{Europhys. Lett.} \textbf{\bibinfo{volume}{13}},
  \bibinfo{pages}{117} (\bibinfo{year}{2008}), \\ \XARXIV{0710.0515},
  \\ \XDOI{10.1088/0143-0807/13/3/003}.

\bibitem[{\citenamefont{McCall}(2008)}]{McCall-2008meta}
\bibinfo{author}{\bibfnamefont{M.~W.} \bibnamefont{McCall}},
  \\ \bibinfo{journal}{Metamaterials} \textbf{\bibinfo{volume}{2}},
  \bibinfo{pages}{92} (\bibinfo{year}{2008}),
  \\ \XDOI{10.1016/j.metmat.2008.05.001}.

\bibitem[{\citenamefont{Hillery and Mlodinow}(1984)}]{Hillery-M-1984pra}
\bibinfo{author}{\bibfnamefont{M.}~\bibnamefont{Hillery}} \bibnamefont{and}
  \bibinfo{author}{\bibfnamefont{L.~D.} \bibnamefont{Mlodinow}},
  \\ \bibinfo{journal}{Phys. Rev. A} \textbf{\bibinfo{volume}{30}},
  \bibinfo{pages}{1860} (\bibinfo{year}{1984}),
  \\ \XDOI{10.1103/PhysRevA.30.1860}.

\bibitem[{\citenamefont{Datta}(1984)}]{Datta-1984ejp}
\bibinfo{author}{\bibfnamefont{S.}~\bibnamefont{Datta}}, \\ \bibinfo{journal}{Eur.
  J. Phys.} \textbf{\bibinfo{volume}{5}}, \bibinfo{pages}{243}
  (\bibinfo{year}{1984}),
  \\ \XWEB{http://www.iop.org/EJ/abstract/0143-0807/5/4/010/}.

\bibitem[{\citenamefont{Carpenter}(1999)}]{Carpenter-1999ieesmt}
\bibinfo{author}{\bibfnamefont{C.~J.} \bibnamefont{Carpenter}},
  \\ \bibinfo{journal}{IEE Proceedings - Science, Measurement and Technology}
  \textbf{\bibinfo{volume}{146}}, \bibinfo{pages}{73} (\bibinfo{year}{1999}),
  \\ \XDOI{10.1049/ip-smt:19990026}.

\bibitem[{\citenamefont{Crenshaw and Akozbek}(2006)}]{Crenshaw-A-2006pre}
\bibinfo{author}{\bibfnamefont{M.~E.} \bibnamefont{Crenshaw}} \bibnamefont{and}
  \bibinfo{author}{\bibfnamefont{N.}~\bibnamefont{Akozbek}},
  \\ \bibinfo{journal}{Phys. Rev. E} \textbf{\bibinfo{volume}{73}},
  \bibinfo{pages}{056613} (\bibinfo{year}{2006}),
  \\ \XWEB{https://doi.org/10.1103/PhysRevE.73.056613}.

\bibitem[{\citenamefont{Campos and Jimenez}(1992)}]{Campos-J-1992ejp}
\bibinfo{author}{\bibfnamefont{I.}~\bibnamefont{Campos}} \bibnamefont{and}
  \bibinfo{author}{\bibfnamefont{J.~L.} \bibnamefont{Jimenez}},
  \\ \bibinfo{journal}{Eur. J. Phys.} \textbf{\bibinfo{volume}{13}},
  \bibinfo{pages}{117} (\bibinfo{year}{1992}),
  \\ \XWEB{http://www.iop.org/EJ/refs/0143-0807/13/3/003}.

\bibitem[{\citenamefont{Obukhov and Hehl}(2003)}]{Obukhov-H-2003pla}
\bibinfo{author}{\bibfnamefont{Y.~N.} \bibnamefont{Obukhov}} \bibnamefont{and}
  \bibinfo{author}{\bibfnamefont{F.~W.} \bibnamefont{Hehl}},
  \\ \bibinfo{journal}{Phys. Lett. A} \textbf{\bibinfo{volume}{311}},
  \bibinfo{pages}{277} (\bibinfo{year}{2003}), 
  \\ \XDOI{10.1016/S0375-9601(03)00503-6}.

\bibitem[{\citenamefont{Raabe and Welsch}(2005)}]{Raabe-W-2005pra}
\bibinfo{author}{\bibfnamefont{C.}~\bibnamefont{Raabe}} \bibnamefont{and}
  \bibinfo{author}{\bibfnamefont{D.-G.} \bibnamefont{Welsch}},
  \\ \bibinfo{journal}{Phys. Rev. A} \textbf{\bibinfo{volume}{71}},
  \bibinfo{pages}{013814} (\bibinfo{year}{2005}), \\ \bibinfo{note}{see comment
  Pitaevskii-2006pra and reply Raabe-W-2006pra; also later comment
  Brevik-E-2008pra.},
  \\ \XDOI{10.1103/PhysRevA.71.013814}.

\bibitem[{\citenamefont{Markel}(2008)}]{Markel-2008oe}
\bibinfo{author}{\bibfnamefont{V.~A.} \bibnamefont{Markel}},
  \\ \bibinfo{journal}{Opt. Express} \textbf{\bibinfo{volume}{16}},
  \bibinfo{pages}{19152} (\bibinfo{year}{2008}), \\ \XARXIV{0712.0605},
  \\ \XDOI{10.1364/OE.16.019152}.

\bibitem[{\citenamefont{Pitaevskii}(2006)}]{Pitaevskii-2006pra}
\bibinfo{author}{\bibfnamefont{L.~P.} \bibnamefont{Pitaevskii}},
  \\ \bibinfo{journal}{Phys. Rev. A} \textbf{\bibinfo{volume}{73}},
  \bibinfo{pages}{047801} (\bibinfo{year}{2006}),
  \\ \XDOI{10.1103/PhysRevA.73.047801}.

\bibitem[{\citenamefont{Raabe and Welsch}(2006)}]{Raabe-W-2006pra}
\bibinfo{author}{\bibfnamefont{C.}~\bibnamefont{Raabe}} \bibnamefont{and}
  \bibinfo{author}{\bibfnamefont{D.-G.} \bibnamefont{Welsch}},
  \\ \bibinfo{journal}{Phys. Rev. A} \textbf{\bibinfo{volume}{73}},
  \bibinfo{pages}{047802} (\bibinfo{year}{2006}),
  \\ \XDOI{10.1103/PhysRevA.73.047802}.

\bibitem[{\citenamefont{Brevik and Ellingsen}(2009)}]{Brevik-E-2008pra}
\bibinfo{author}{\bibfnamefont{I.}~\bibnamefont{Brevik}} \bibnamefont{and}
  \bibinfo{author}{\bibfnamefont{S.~A.} \bibnamefont{Ellingsen}},
  \\ \bibinfo{journal}{Phys. Rev. A} \textbf{\bibinfo{volume}{79}},
  \bibinfo{pages}{027801} (\bibinfo{year}{2009}),
  \\ \XDOI{10.1103/PhysRevA.79.027801}.

\bibitem[{\citenamefont{Buchwald}(1985)}]{Buchwald-FMTM}
\bibinfo{author}{\bibfnamefont{J.~Z.} \bibnamefont{Buchwald}},
  \emph{\bibinfo{title}{From Maxwell to microphysics}}
  (\bibinfo{publisher}{University of Chicago}, \bibinfo{year}{1985}), ISBN
  \bibinfo{isbn}{0-226-07883-3}.

\bibitem[{\citenamefont{Kolesik et~al.}(2002)\citenamefont{Kolesik, Moloney,
  and Mlejnek}}]{Kolesik-MM-2002prl}
\bibinfo{author}{\bibfnamefont{M.}~\bibnamefont{Kolesik}},
  \bibinfo{author}{\bibfnamefont{J.~V.} \bibnamefont{Moloney}},
  \bibnamefont{and} \bibinfo{author}{\bibfnamefont{M.}~\bibnamefont{Mlejnek}},
  \\ \bibinfo{journal}{Phys. Rev. Lett.} \textbf{\bibinfo{volume}{89}},
  \bibinfo{pages}{283902} (\bibinfo{year}{2002}),
  \\ \XDOI{10.1103/PhysRevLett.89.283902}.

\bibitem[{\citenamefont{Kolesik and Moloney}(2004)}]{Kolesik-M-2004pre}
\bibinfo{author}{\bibfnamefont{M.}~\bibnamefont{Kolesik}} \bibnamefont{and}
  \bibinfo{author}{\bibfnamefont{J.~V.} \bibnamefont{Moloney}},
  \\ \bibinfo{journal}{Phys. Rev. E} \textbf{\bibinfo{volume}{70}},
  \bibinfo{pages}{036604} (\bibinfo{year}{2004}),
  \\ \XDOI{10.1103/PhysRevE.70.036604}.

\bibitem[{\citenamefont{Kinsler et~al.}(2005)\citenamefont{Kinsler, Radnor, and
  New}}]{Kinsler-RN-2005pra}
\bibinfo{author}{\bibfnamefont{P.}~\bibnamefont{Kinsler}},
  \bibinfo{author}{\bibfnamefont{S.~B.~P.} \bibnamefont{Radnor}},
  \bibnamefont{and} \bibinfo{author}{\bibfnamefont{G.~H.~C.}
  \bibnamefont{New}}, \\ \bibinfo{journal}{Phys. Rev. A}
  \textbf{\bibinfo{volume}{72}}, \bibinfo{pages}{063807}
  (\bibinfo{year}{2005}), \\ \bibinfo{note}{note that in this reference, the
  convolution symbol between the $\alpha_c \beta_r$ coefficents and the $G^\pm$
  terms in square brackets in the frequency-domain propagation equations was
  inadvertently omitted}, \\ \XARXIV{physics/0611215v1},
  \\ \XDOI{10.1103/PhysRevE.75.066603}.

\bibitem[{\citenamefont{Kinsler}(2006)}]{Kinsler-2006arXiv-fleck}
\bibinfo{author}{\bibfnamefont{P.}~\bibnamefont{Kinsler}}
  (\bibinfo{year}{2006}), \\ \bibinfo{note}{``Theory of directional pulse
  propagation: detailed calculations''},
  \\ \XARXIV{physics/0611216},
  \\ \XWEB{http://arxiv.org/abs/physics/0611216}.

\bibitem[{\citenamefont{Mizuta et~al.}(2005)\citenamefont{Mizuta, Nagasawa,
  Ohtani, and Yamashita}}]{Mizuta-NOY-2005pra}
\bibinfo{author}{\bibfnamefont{Y.}~\bibnamefont{Mizuta}},
  \bibinfo{author}{\bibfnamefont{M.}~\bibnamefont{Nagasawa}},
  \bibinfo{author}{\bibfnamefont{M.}~\bibnamefont{Ohtani}}, \bibnamefont{and}
  \bibinfo{author}{\bibfnamefont{M.}~\bibnamefont{Yamashita}},
  \\ \bibinfo{journal}{Phys. Rev. A} \textbf{\bibinfo{volume}{72}},
  \bibinfo{pages}{063802} (\bibinfo{year}{2005}),
  \\ \XDOI{10.1103/PhysRevA.72.063802}.

\bibitem[{\citenamefont{Favaro et~al.}(2009)\citenamefont{Favaro, Kinsler, and
  McCall}}]{Favaro-KM-2009oe}
\bibinfo{author}{\bibfnamefont{A.}~\bibnamefont{Favaro}},
  \bibinfo{author}{\bibfnamefont{P.}~\bibnamefont{Kinsler}}, \bibnamefont{and}
  \bibinfo{author}{\bibfnamefont{M.~W.} \bibnamefont{McCall}},
  \\ \bibinfo{journal}{Opt. Express} \textbf{\bibinfo{volume}{17}},
  \bibinfo{pages}{15167} (\bibinfo{year}{2009}),  \\ 
  \XDOI{10.1364/OE.17.015167}.

\bibitem[{\citenamefont{Boyd}(2008)}]{Boyd-NLO}
\bibinfo{author}{\bibfnamefont{R.~W.} \bibnamefont{Boyd}},
  \emph{\bibinfo{title}{Nonlinear Optics}} (\bibinfo{publisher}{Academic Press
  Inc.}, \bibinfo{address}{New York}, \bibinfo{year}{2008}),
  \bibinfo{edition}{3rd} ed., ISBN \bibinfo{isbn}{978-0-12-369470-6},
  \\ \bibinfo{note}{1st ed. 1994, 2nd ed. 2003}.

\bibitem[{\citenamefont{Brabec and Krausz}(1997)}]{Brabec-K-1997prl}
\bibinfo{author}{\bibfnamefont{T.}~\bibnamefont{Brabec}} \bibnamefont{and}
  \bibinfo{author}{\bibfnamefont{F.}~\bibnamefont{Krausz}},
  \\ \bibinfo{journal}{Phys. Rev. Lett.} \textbf{\bibinfo{volume}{78}},
  \bibinfo{pages}{3282} (\bibinfo{year}{1997}),
  \\ \XDOI{10.1103/PhysRevLett.78.3282}.

\bibitem[{\citenamefont{Geissler et~al.}(1999)\citenamefont{Geissler, Tempea,
  Scrinzi, Schn{\"u}rer, Krausz, and Brabec}}]{Geissler-TSSKB-1999prl}
\bibinfo{author}{\bibfnamefont{M.}~\bibnamefont{Geissler}},
  \bibinfo{author}{\bibfnamefont{G.}~\bibnamefont{Tempea}},
  \bibinfo{author}{\bibfnamefont{A.}~\bibnamefont{Scrinzi}},
  \bibinfo{author}{\bibfnamefont{M.}~\bibnamefont{Schn{\"u}rer}},
  \bibinfo{author}{\bibfnamefont{F.}~\bibnamefont{Krausz}}, \bibnamefont{and}
  \bibinfo{author}{\bibfnamefont{T.}~\bibnamefont{Brabec}},
  \\ \bibinfo{journal}{Phys. Rev. Lett.} \textbf{\bibinfo{volume}{83}},
  \bibinfo{pages}{2930} (\bibinfo{year}{1999}),
  \\ \XDOI{10.1103/PhysRevLett.83.2930}.

\bibitem[{\citenamefont{Kinsler and New}(2003)}]{Kinsler-N-2003pra}
\bibinfo{author}{\bibfnamefont{P.}~\bibnamefont{Kinsler}} \bibnamefont{and}
  \bibinfo{author}{\bibfnamefont{G.~H.~C.} \bibnamefont{New}},
  \\ \bibinfo{journal}{Phys. Rev. A} \textbf{\bibinfo{volume}{67}},
  \bibinfo{pages}{023813} (\bibinfo{year}{2003}), \\ \XARXIV{physics/0212016v1},
  \\ \XDOI{10.1103/PhysRevA.67.023813}.

\bibitem[{\citenamefont{Kinsler}(2007{\natexlab{a}})}]{Kinsler-2007-envel}
\bibinfo{author}{\bibfnamefont{P.}~\bibnamefont{Kinsler}}
  (\bibinfo{year}{2007}{\natexlab{a}}), \\ \bibinfo{note}{``Pulse propagation
  methods in nonlinear optics''}, \\ \XARXIV{0707.0982},
  \\ \XWEB{http://arxiv.org/abs/0707.0982}.

\bibitem[{\citenamefont{Genty et~al.}(2007)\citenamefont{Genty, Kinsler,
  Kibler, and Dudley}}]{Genty-KKD-2007oe}
\bibinfo{author}{\bibfnamefont{G.}~\bibnamefont{Genty}},
  \bibinfo{author}{\bibfnamefont{P.}~\bibnamefont{Kinsler}},
  \bibinfo{author}{\bibfnamefont{B.}~\bibnamefont{Kibler}}, \bibnamefont{and}
  \bibinfo{author}{\bibfnamefont{J.~M.} \bibnamefont{Dudley}},
  \\ \bibinfo{journal}{Opt. Express} \textbf{\bibinfo{volume}{15}},
  \bibinfo{pages}{5382} (\bibinfo{year}{2007}),
  \\ \XDOI{10.1364/OE.15.005382}.

\bibitem[{\citenamefont{Kinsler}(2007{\natexlab{b}})}]{Kinsler-2007josab}
\bibinfo{author}{\bibfnamefont{P.}~\bibnamefont{Kinsler}}, \\ \bibinfo{journal}{J.
  Opt. Soc. Am. B} \textbf{\bibinfo{volume}{24}}, \bibinfo{pages}{2363}
  (\bibinfo{year}{2007}{\natexlab{b}}), \\ \bibinfo{note}{the arXiv:0707.0986
  version contains an additional appendix.}, \\ \XARXIV{0707.0986v2},
  \\ \XDOI{10.1364/JOSAB.24.002363}.

\bibitem[{\citenamefont{Kinsler}(2008)}]{Kinsler-2008-fbdiff}
\bibinfo{author}{\bibfnamefont{P.}~\bibnamefont{Kinsler}}
  (\bibinfo{year}{2008}), \\ \bibinfo{note}{``Transverse limits to the
  uni-directional pulse propagation approximation''}, \\ \XARXIV{0810.5701},
  \\ \XWEB{http://arxiv.org/abs/0810.5701}.

\bibitem[{\citenamefont{Kinsler}(2010)}]{Kinsler-2010pra-fchhg}
\bibinfo{author}{\bibfnamefont{P.}~\bibnamefont{Kinsler}},
  \\ \bibinfo{journal}{Phys. Rev. A} \textbf{\bibinfo{volume}{81}},
  \bibinfo{pages}{013819} (\bibinfo{year}{2010}), \\ \XARXIV{0810.5689},
  \\ \XDOI{10.1103/PhysRevA.81.013819}.

\bibitem[{\citenamefont{Kinsler}(2011)}]{Kinsler-2011ejp}
\bibinfo{author}{\bibfnamefont{P.}~\bibnamefont{Kinsler}},
  \\ \bibinfo{journal}{Eur. J. Phys.} \textbf{\bibinfo{volume}{32}},
  \bibinfo{pages}{1687} (\bibinfo{year}{2011}), \\ \bibinfo{note}{the arXiv
  version has additional appendices}, \\ \XARXIV{1106.1792},
  \\ \XDOI{10.1088/0143-0807/32/6/022}.

\bibitem[{\citenamefont{Kinsler}(2015)}]{Kinsler-2015arxiv-d2owe}
\bibinfo{author}{\bibfnamefont{P.}~\bibnamefont{Kinsler}}
  (\bibinfo{year}{2015}), \\ \bibinfo{note}{``Temporally propagated optical
  pulses, and what they reveal about dispersion handling''},
  \\ \XARXIV{1501.05569}, \\ \XWEB{http://arxiv.org/abs/1501.05569}.

\bibitem[{\citenamefont{Yee}(1966)}]{Yee-1966tap}
\bibinfo{author}{\bibfnamefont{K.~S.} \bibnamefont{Yee}},
  \\ \bibinfo{journal}{IEEE Trans. Antennas Propagat.}
  \textbf{\bibinfo{volume}{14}}, \bibinfo{pages}{302} (\bibinfo{year}{1966}),
  \\ \XDOI{10.1109/TAP.1966.1138693}.

\bibitem[{\citenamefont{Scalora and Crenshaw}(1994)}]{Scalora-C-1994oc}
\bibinfo{author}{\bibfnamefont{M.}~\bibnamefont{Scalora}} \bibnamefont{and}
  \bibinfo{author}{\bibfnamefont{M.~E.} \bibnamefont{Crenshaw}},
  \\ \bibinfo{journal}{Opt. Comm.} \textbf{\bibinfo{volume}{108}},
  \bibinfo{pages}{191} (\bibinfo{year}{1994}),
  \\ \XDOI{10.1016/0030-4018(94)90647-5}.

\bibitem[{\citenamefont{Authors}(2003)}]{FocusIssue-2003oe-nrm}
\bibinfo{author}{\bibfnamefont{Various}~\bibnamefont{authors}},
  \\ \bibinfo{journal}{Opt. Express} \textbf{\bibinfo{volume}{11}},
  \bibinfo{pages}{639} (\bibinfo{year}{2003}), \\ \bibinfo{note}{"Focus Issue:
  Negative refraction and metamaterials"},
  \\ \XWEB{http://www.opticsinfobase.org/oe/issue.cfm?volume=11&issue=7}.

\bibitem[{\citenamefont{Brabec and Krausz}(2000)}]{Brabec-K-2000rmp}
\bibinfo{author}{\bibfnamefont{T.}~\bibnamefont{Brabec}} \bibnamefont{and}
  \bibinfo{author}{\bibfnamefont{F.}~\bibnamefont{Krausz}},
  \\ \bibinfo{journal}{Rev. Mod. Phys.} \textbf{\bibinfo{volume}{72}},
  \bibinfo{pages}{545} (\bibinfo{year}{2000}),
  \\ \XDOI{10.1103/RevModPhys.72.545}.

\end{thebibliography}

%

\end{document}